\documentclass[twocolumn]{webofc}

\usepackage[varg]{txfonts}   

\def\be{\begin{eqnarray} &&}

\def\ee{\end{eqnarray}}
 
\def\sla#1{\rlap\slash #1}

\usepackage{graphicx} 
\usepackage{dcolumn} 
\usepackage{bm} 

\usepackage{epsfig}

\woctitle{MENU-2013}
\begin{document}
\title{ Spin-1 Particles with Light-Front Approach  }

\author{
J.~P.~B.~C.~de~Melo \inst{1}
\fnsep\thanks{\email{joao.mello@cruzeirodosul.edu.br}} 
\and
    Anac\'e~N.~da~Silva\inst{1}
\and
Clayton S.~Mello \inst{1,2}
\and 
  Tobias Frederico  \inst{2} 
}
\institute{Laborat\'orio de F\'\i sica Te\'orica e Computacional,~LFTC, 
Universidade Cruzeiro do Sul, \newline 
S\~ao Paulo,~Brazil,~01506-000 
\and
Departamento de F\'\i sica, Instituto  Tecnol\'ogico de Aeron\'autica,
Centro T\'ecnico Aeroespacial,
\newline 
12.228-900,~S\~ao Jos\'e dos Campos, S\~ao Paulo,~Brazil
                     }
\abstract{
For the vector sector, i.e, mesons with spin-1,
 the electromagnetic form factors and
 anothers observables are calculated with the light-front approach.
 However, the light-front quantum field theory
 have some problems, for example, the rotational
 symmetry breaking. We solve that problem added the zero modes contribuition to the
 matrix elements of the electromagnetic current, besides the valence contribuition.
   We found that among the four independent matrix
 elements of the plus component in the light-front helicity basis
 only the $0\to 0$ one carries zero mode contributions.
}
\maketitle

\section{Introduction}
\label{intro}
The main objective of the {\it QCD}, (quantum cromodynamics) quantum field theory is explain the 
bound state hadronic structure; mesons and baryons, in terms of the 
fundamental block of the matter,ie., quarks and gluons.
However, it is no simple task in the 
quantum field theory~\cite{Muta}.  
By another side, the strong interactions of the hadrons, 
is not described by the perturbative quantum field theory. The alternative 
is the light-front quantum field theory~({\it LFQFT}),  the 
natural approach to hadronic bound state systems~\cite{Brodsky98}, 
the first time,  formulated by Paul Dirac some years ago~\cite{Dirac1949}. 
The spin-1 particles have an more complicate structure with compared hadronic 
spin zero structure, for example the pion 
meson and 
kaon~\cite{Pacheco99,Maris97,Pacheco2002,Pacheco2006,
Fabiano2007,Leitner2011,Edson2012}. 
However in the last years, many works dedicate to the structure of spin-1 
particles appears with differents 
approachs~(see the references~\cite{Pacheco97,Hawes1999,Tomasi2007,Wang2007,Braguta2010,
Tobias2010,Choi2011,Maris2008,Pacheco2012,Roberts2013}). 
In this work, we use an light-front constituent quark 
model~({\it LFCQM})~\cite{Pacheco97} in 
order to calculate some observables for spin-1 particle, i.e, the rho meson. 
However, with the light-front approach, besides the 
valence components of the electromagnetic current, in order to keep the 
covariance, is necessary to add the non-valence components of the 
electromagnetic current~\cite{Pacheco97,Pacheco99}.

\section{The Model}
The matrix elements of the electromagnetic current 
${\cal J}_{ji}={\epsilon^\prime_j}^\alpha \epsilon^\beta_iJ_{\alpha
\beta}^{\mu}$ for spin-1 particles, 
in the impulse approximation are written as~\cite{Pacheco97}, 
for the plus component of the  matrix 
elements of the electromagnetic current:
\begin{eqnarray}
J^+_{ji}&=&\imath  \int \frac{d^4k}{(2\pi)^4}
 \frac{ Tr[ 
 \epsilon^{'\nu}_j \Gamma_{\beta}(k,k-p_f)
(\sla{k}-\sla{p_f} +m) \gamma^{+} }
{((k-p_i)^2 - m^2+\imath\epsilon) 
(k^2 - m^2+\imath \epsilon)}
\nonumber \\
& & \frac{
(\sla{k}-\sla{p_i}+m) \epsilon^\mu_i \Gamma_{\alpha}(k,k-p_i)(\sla{k}+m) ] }
{((k-p_f)^2 - m^2+\imath \epsilon) }
\nonumber \\ & &  
\times \Lambda(k,p_f)\Lambda(k,p_i) \ .
\label{eq:tria}
\end{eqnarray}
where $\epsilon^{\prime \nu}_j$ and $\epsilon^{\mu}_i$ are the
polarization four-vectors of the final and initial states,
respectively and $m$ is the constituent quark mass. In the equation above, 
the~$\Lambda(k,p)=1/((k-p)^2 -m^2_R + \imath \epsilon )$, 
is the regulator function in order to turn the Feynman amplitude finite.
The explicity polarizations used here, are 
$\epsilon^{\mu}_{x}=(-\sqrt{\eta},\sqrt{1+\eta},0,0),
\epsilon_y^{\mu}=(0,0,1,0),\epsilon_z^{\mu}= (0,0,0,1)$,~for the initial state and the final state, 
$\epsilon^{'\mu}_{x}=(\sqrt{\eta},\sqrt{1+\eta},0,0), 
\epsilon_y^{'\mu}=(0,0,1,0),\epsilon_z^{'\mu}=(0,0,0,1)$. 
The vertex model to the bound state spin-1 particle and pair~$q\bar{q}$ 
utilized here is~\cite{Pacheco97}:
\begin{equation}
\Gamma^\mu (k,p) = \gamma^\mu -\frac{m_v}{2}
\frac{2 k^\mu -p^\mu}
{ p.k + m_{v} m -\imath \epsilon}   \ ,
\label{rhov}
\end{equation}
the vector particle is on-shell and $m_v$ is the vector bound state mass. 
After the integration in the light-front energy, $k^-=\frac{\vec{k}^2_{\perp}+ m^2}{k^+}$, 
the light-front valence wave function is obtained: 
\begin{equation}
\Phi_i(x,\vec k_\perp)=
\frac{N^2}{(1-x)^2(m^2_v - M_0^2)
(m^2_v - M^2_R)^2} 
\vec \epsilon_i . [\vec \gamma -  \frac{\vec k}
{\frac{M_0}{2}+ m}]  \ . 
\label{eq:npwf}
\end{equation} 

\section{Electromagnetic Form Factors}

The electromagnetic form factors for spin-1 particles, 
are linear combinations of the electromagnetic 
current~\cite{Pacheco97,Cardarelli95}. However, 
for spin-1 particles, we have 
four matrix elements and only three electromagnetic form factors, 
$G_0,G_1$ and $G_2$, ie., charge, magnetic and quadrupole electromagnetic form 
factors respectively, 
then, the combinations of the matrix elements to extract the 
electromagnetic form factors, present some arbitrary in the {\it LFQFT}. 

But, in the {\it LFQFT}, thanks by the 
the angular condition~\cite{Cardarelli95,Inna84}, 
\begin{eqnarray}
\Delta(q^2) & = & 
(1+2 \eta)
I^{+}_{11}+I^{+}_{1-1} - \sqrt{8 \eta}
I^{+}_{10} - I^{+}_{00} \nonumber \\ 
& = & \left( J^+_{yy} - J^+_{zz} \right)
\left( 1+ \eta \right) ~,
\label{angular}
\end{eqnarray}
it is possible eliminate some matrix elements of the 
electromagnetic current and extract the electromagnetic form factors.
In the literature, we have four prescriptions to 
combine the matrix elements of the electromagnetic current in order to 
calculate the electromagnetic form factors~\cite{Pacheco97,Cardarelli95}, 
using the valence part of the electromagnetic current.

In this work, we use two 
availables prescriptions in the literature
~\cite{Inna84,Hiller92} and compare that 
prescriptions with the instant form basis. In the case, of the reference~\cite{Inna84}, 
the elimination of matrix elements $I^+_{00}$, produce the 
following expression to the electromagnetic form factors:
 \begin{eqnarray}
G_0^{GK}&=&\frac{1}{3}[(3-2 \eta) I^{+}_{11}+
2 \sqrt{2 \eta} I^{+}_{10} +  I^{+}_{1-1}] \nonumber \\  
& = & \frac{1}{3}[J_{xx}^{+}+ 2 J_{yy}^{+}-\eta  J_{yy}^{+} 
+ \eta  J_{zz}^{+}]~,
\nonumber \\
G_1^{GK}&=&2 [I^{+}_{11}-\frac{1}{ \sqrt{2 \eta}} I^{+}_{10}] \nonumber \\
&  = & J_{yy}^{+} -  ( J_{zz}^{+}+\frac{J_{zx}^{+}}{\sqrt{\eta}} )~,
\nonumber \\
G_2^{GK}&=&\frac{2 \sqrt{2}}{3}[- \eta I^{+}_{11}+
\sqrt{2 \eta} I^{+}_{10} -  I^{+}_{1-1}] \nonumber \\
& = & \frac{\sqrt{2}}{3}[J_{xx}^{+}+J_{yy}^{+} (-1-\eta)
+\eta  J_{zz}^{+}].
\label{inna}
\end{eqnarray}
In the reference~\cite{Hiller92}, the $I^+_{11}$ 
matrix element of the electromagnetic current is eliminate from 
Eq.~(\ref{angular}), 
and the eletromagnetic form factors to spin-1 particles are 
given by:
\begin{eqnarray}
G_0^{BH}&=&\frac{1}{3(1+\eta)}[(3-2 \eta) I^{+}_{00}+
8 \sqrt{2 \eta} I^{+}_{10} +  \nonumber \\
& & 2 (2 \eta -1) I^{+}_{1-1}] \nonumber \\
& = &\frac{1}{3 (1+2 \eta)}[J_{xx}^{+} (1+2 \eta) + 
J_{yy}^{+}(2 \eta-1)  \nonumber \\
& & + J_{zz}^{+}(3+2 \eta)]~,
\nonumber \\
G_1^{BH}&=&\frac{2}{(1+2 \eta)}[I^{+}_{00}
-I^{+}_{1-1}+\frac{(2 \eta -1)}{\sqrt{2 \eta}} I^{+}_{10}]
\nonumber \\
&=&\frac{1}{(1+2 \eta)}[\frac{J_{zx}^{+}}{\sqrt{\eta}}
 (1+2 \eta)- J_{yy}^{+} +  J_{zz}^{+}]~,
\nonumber \\
G_2^{BH} & = & \frac{2 \sqrt{2}}{3 (1+ 2 \eta)}[\sqrt{2 \eta} I^{+}_{10}
-\eta I^{+}_{00} -( \eta+1) I^{+}_{1-1}]
\nonumber \\
&= & \frac{ \sqrt{2}}{3 (1+2 \eta)}
[ J_{xx}^{+} (1+2 \eta)-  J_{yy}^{+}(1+ \eta) 
- \eta J_{zz}^{+}]~.
\nonumber \\
\label{PREBH}
\end{eqnarray}
Was shown in the references~\cite{Pacheco99,Pacheco97,Ji2002,Choi2004}, 
if the zero modes or pair terms contribuitions are ignored, the rotational symmetry 
is breaking and the covariance are lost. After the inclusion of the zero modes, 
the full covariance of the current electromagnetic current is restored.

In the references~\cite{Pacheco2004,Pacheco2012}, are 
found some relations 
between the matrix elements of the plus component of 
electromagnetic current for the Z-terms contribuitions of 
the electromagnetic current:
\begin{equation}
J^{+Z}_{xx} +\eta \  J^{+Z}_{zz}=0 \ ,    J^{+Z}_{zx} +\sqrt{\eta} \
J^{+Z}_{zz}=0\ ~~\text{and}~~ J^{+Z}_{yy}=0 . 
\label{finalcurr}
\end{equation}
The relation above, takes the following relations for the electromagnetic matrix 
elements for the valence components of the current whitout Z-terms:
(the superscript $V$
indicates the valence terms):   
\begin{eqnarray} 
J^{+}_{xx} & = & J^{+V}_{xx}-\eta \left(J^{+V}_{yy}-J^{+V}_{zz}\right),  \nonumber \\
J^{+}_{zx} & = & J^{+V}_{zx}-\sqrt{\eta}\left(J^{+V}_{yy}-J^{+V}_{zz}\right),  \nonumber  \\
J^+_{zz}   & = & J^{+V}_{yy} , \nonumber \\
J^{+Z}_{zz}& = & J^{+V}_{yy} - J^{+V}_{zz} \ .
\label{zrcurr} 
\end{eqnarray}
Using the above relationships in the Eq.(\ref{PREBH}), we obtain
following expressions for the electromagnetic form factors 
of the spin-1 particles,
\begin{eqnarray}
G_0^{BH}&=& \frac{1}{3 (1+2 \eta)}[J_{xx}^{+} (1+2 \eta)+ J_{yy}^{+}(2 \eta-1) 
+ J_{zz}^{+}(3+2 \eta)] \nonumber \\
& = & \frac{1}{3}[J^{+V}_{xx} + (2-\eta )J^{+V}_{yy} + \eta J^{+V}_{zz}],
\nonumber \\
G_1^{BH}&=& \frac{1}{(1+2 \eta)}[\frac{J_{zx}^{+}}{\sqrt{\eta}}
 (1+2 \eta)- J_{yy}^{+} +  J_{zz}^{+}]
 \nonumber \\
 & = & [J^{+V}_{yy}- \frac{J^{+V}_{zx}}{\sqrt{\eta}} - J^{+V}_{zz}],
\nonumber \\
G_2^{BH}&=&
\frac{ \sqrt{2}}{3 (1+2 \eta)}
[J_{xx}^{+} (1+2 \eta)- J_{yy}^{+}(1+ \eta) - \eta J_{zz}^{+}]
\nonumber \\
& = & \frac{\sqrt{2}}{3}[ J^{+V}_{xx}-(1+\eta )J^{+V}_{yy} + \eta J^{+V}_{zz}]~,
 \end{eqnarray}
the final expressions for the electromagnetic 
form factors,~$G^{CCKP}_0,G^{CCKP}_1$ and $G^{CCKP}_2$, given exactly the same 
expressions as Grach et al.~\cite{Inna84} and the numerical calculation 
produce the same results compared with the 
instant form approach~\cite{Pacheco2012}~(see the figures).

As shown before, in the references~\cite{Pacheco2004,Choi2004}
the elimination of the matrix element $I^{+}_{00}$, 
by the Inna Grach et al.~\cite{Inna84}, eliminates the zero modes, thus the
result produced for the light-front approach calculations, 
same like the instant form formalism. In the table-~\ref{tab1}, the 
present work are compared with anothers approaches in the 
literature.

\newpage 

\begin{table}
\caption{The rho meson observables are calculated with the light-front quark model 
and compared with other model calculations. 
The units are,~$(fm^2)$ for the radius,~$(e/2m_V)$ for the magnetic moment,
~$(e/m_V^2)$ for the quadrupole and ($MeV$) the decay constant.
The paramters are $m_q=0.430~GeV$,~$m_{\rho}=0.770~GeV$ and $m_R=3.0~GeV$.}
\label{tab1}
\begin{center}
\begin{tabular}{|l|l|l|l|l|l|}
\hline 
\hline
                     &~$f_{\rho}$ &~$<r_{\rho}^2>$ &~$\mu$&$Q_2$    \\
\hline
This work            & ~153    &~0.267  &~2.10  &~0.898     \\
\cite{Hawes1999}     &         & ~0.310 &~2.69  &~0.840    \\
\cite{Maris2008}     & ~221    &~0.540  &~2.01  & ~0.410 \\
 \cite{Roberts2013}  & ~129    &        &~2.11  &  \\
 \cite{Ji2002}       & ~134    &~0.296  &~2.10  &~0.910 \\
\hline
\hline
\end{tabular}
\end{center}
  \end{table}

\begin{figure}
\centering
\sidecaption
\includegraphics[width=8cm,clip]{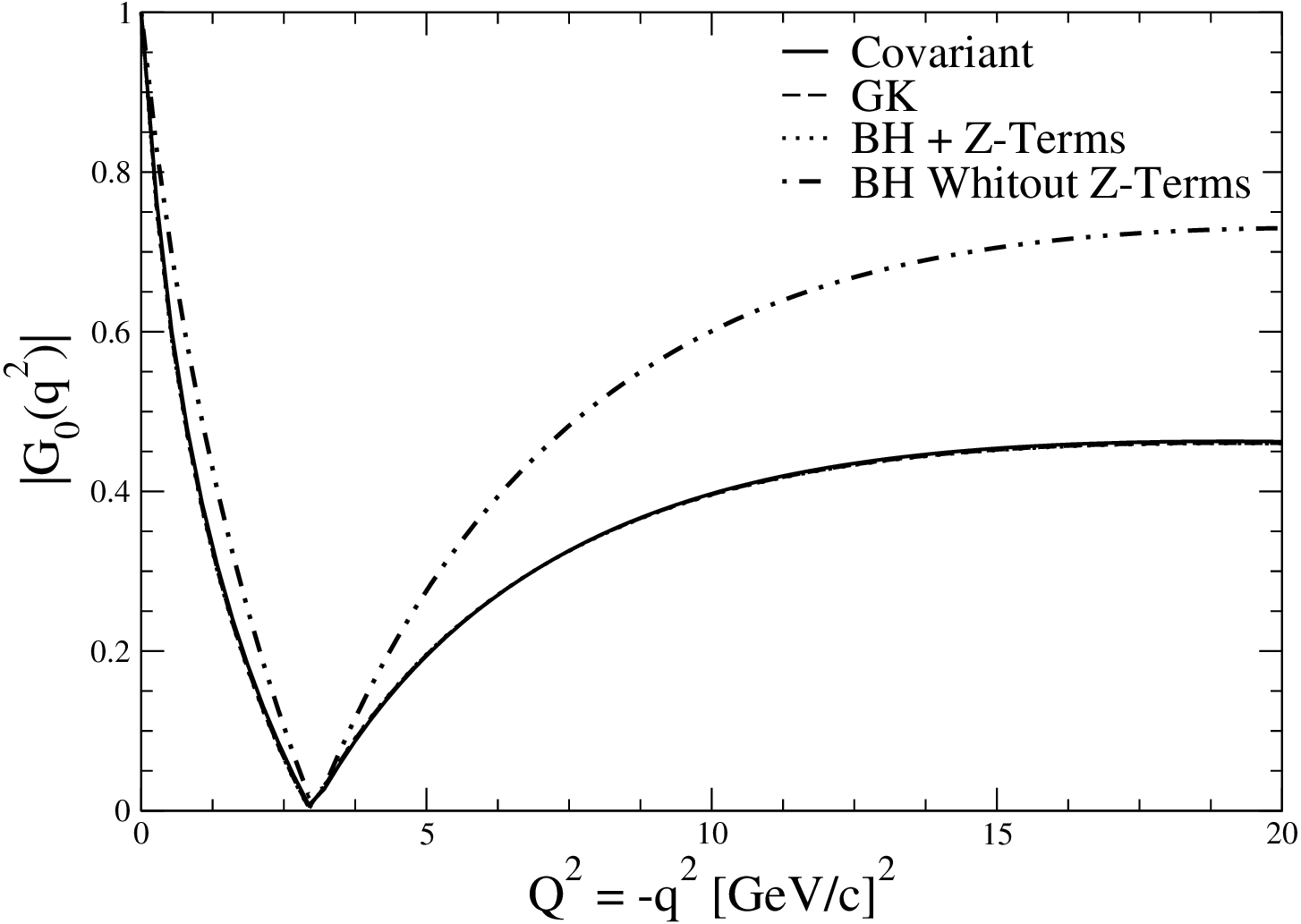}
\caption{Electromagnetic charge form factor, $G_0$, for 
two prescriptions~\cite{Inna84,Hiller92} and compared with the instant form 
calculation. The prescriptions by the authors of the reference 
~\cite{Hiller92}, after the inclusion of the zero modes contribuition, is 
the same of the covariant calculation and the prescriptions in the 
references~\cite{Inna84}.}
\label{fig1} 
\end{figure}

\begin{figure}
\centering
\sidecaption
\includegraphics[width=8cm,clip]{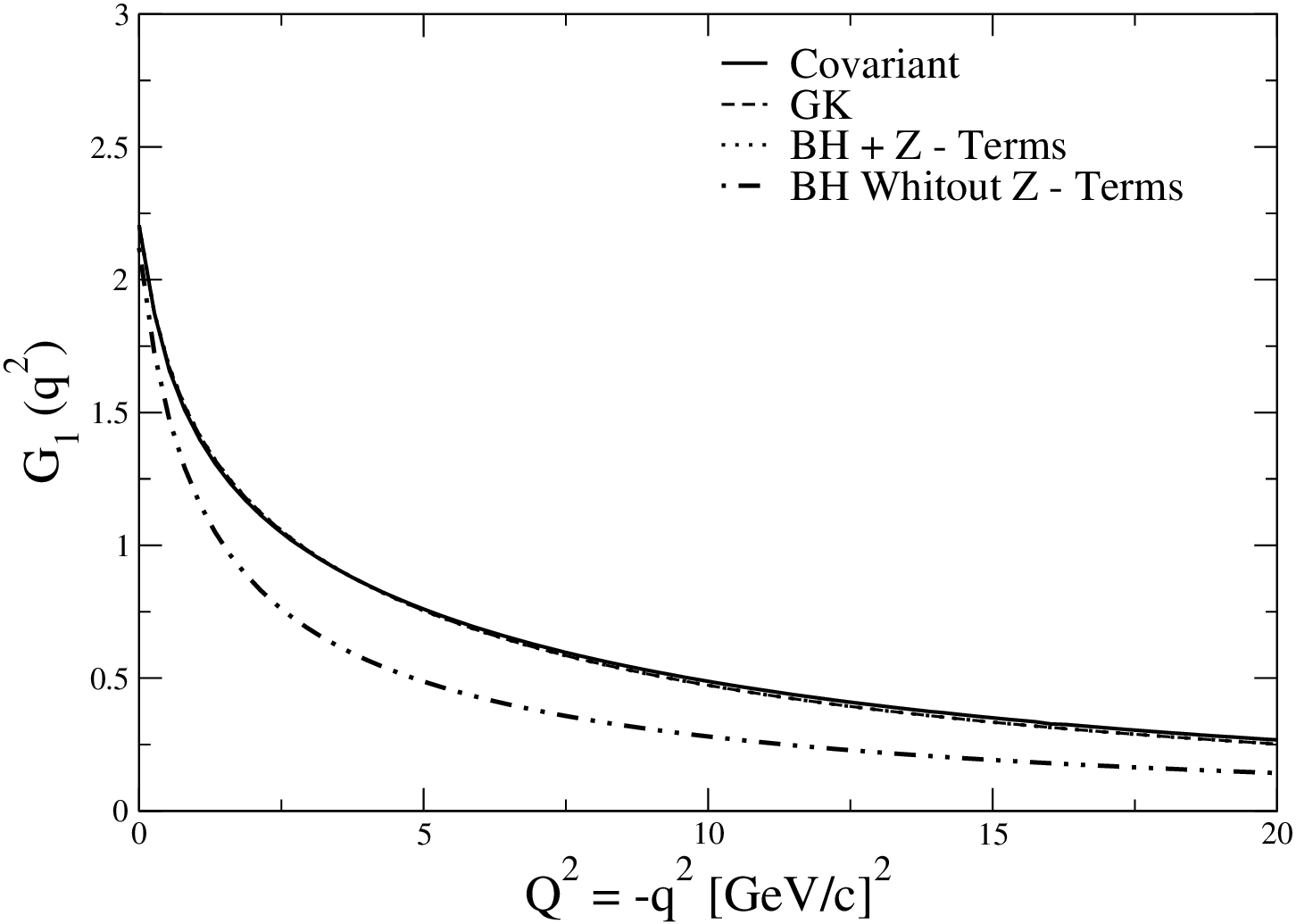}
\caption{Magnetic form factor, $G_1$, 
the labels are the same of the figue 1.}
\label{fig2}       
\end{figure}

\newpage

\section{Conclusion}

In the case of spin-1 particles, terms of no-valence becomes 
necessary to added~\cite {Pacheco2012, Pacheco97, Ji2002, Choi2004}.
However, only term of the matrix elements of the 
electromagnetic current in the basis of the spin on the light-front~$I^+_{00}$, contributes to
the zero modes~\cite {Pacheco2012, Choi2004}.

In this work, we use the relationships between some 
matrix elements of the electromagnetic current of the 
plus component,~"+" for the rho meson~\cite{Pacheco2012}, in order  to
get the electromagnetic form factors free of the zero modes.

We compared two
existing prescriptions in the literature~\cite {Inna84,Hiller92} 
and the calculations are compared with instant form formalism.

\begin{figure}
\centering
\sidecaption
\includegraphics[width=8cm,clip]{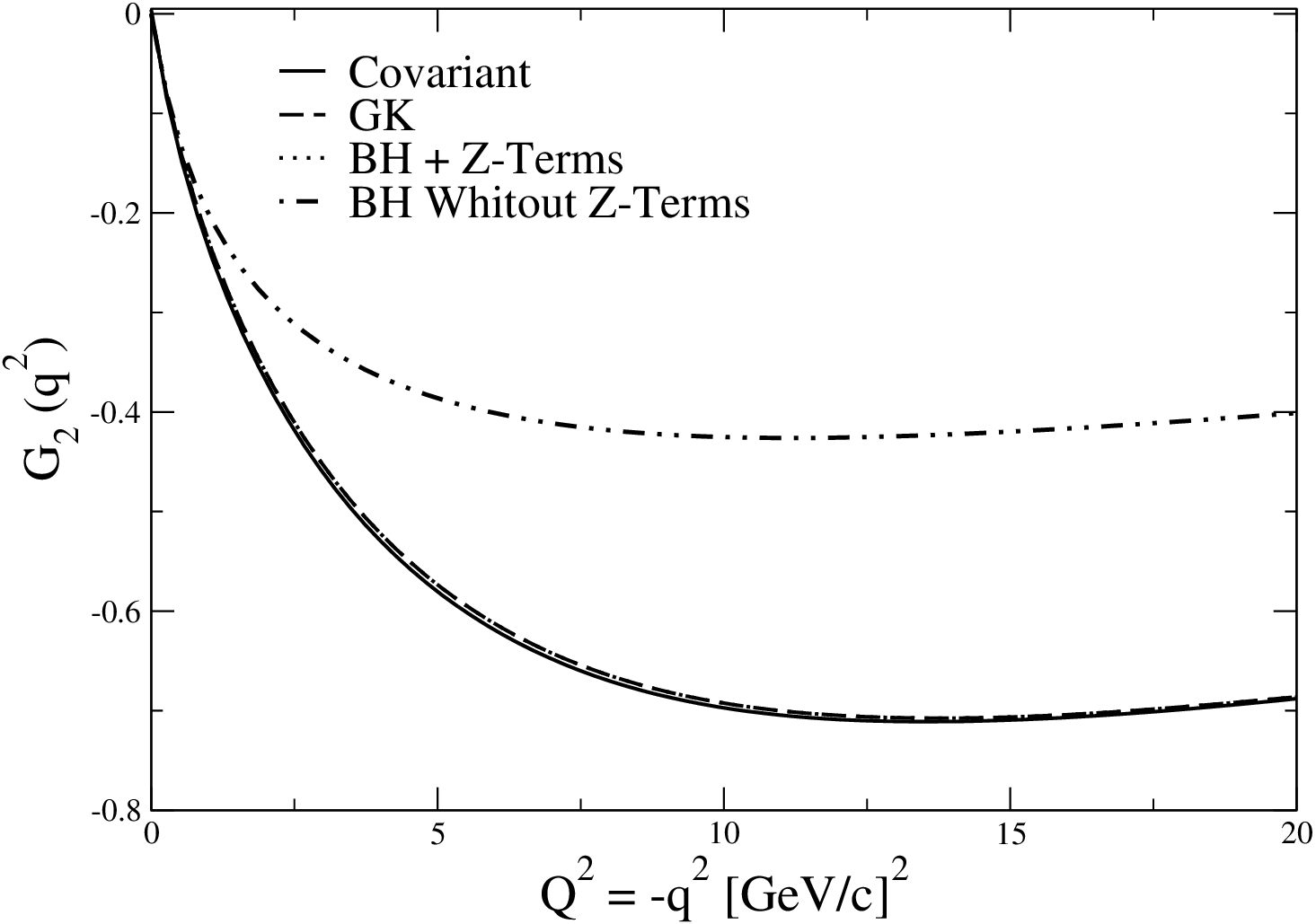}
\caption{Quadrupole form factor, $G_2$, same labels like figure 1.}
\label{fig3}
\end{figure}

{\bf Acknowledgements:}
J.~Pacheco~B.~C.~de~Melo thanks to the organizers MENU2013, Rome, Italy, 
for 
the invitation. This work was supported by the Brazilian
agencies FAPESP (Funda\c{c}\~ao de Amparo a Pesquisa do Estado de
S\~ao Paulo),~CNPq (Conselho Nacional de Desenvolvimento
Ci\^entifico e T\'ecnologico) and CAPES~(
Coordena\c{c}\~ao de Aperfei\c{c}oamento de Pessoal de N\'\i vel Superior).


\begin{thebibliography}{}

\bibitem{Muta} Taizo Muta, \textit{Foundations of Quantum Cromodynamics, 
An Introduction to Perturbative Methods in Gauge Theories}
~(Word Scientific, Singapore,~1987).  

\bibitem{Brodsky98} S.~Brodsky, Hans-Cristian Pauli, Stephen Pinsky, 
Physics Reports~\textbf{301},~299~(1998).

\bibitem{Dirac1949} P.~A.~M.~Dirac,~Rev.~Mod.~Physics~\textbf{21},~392~(1949).

\bibitem{Maris97} P.~Maris,~C.~Roberts,~Phys.~Rev.~{\bf C56}, 3369~(1997).

\bibitem{Pacheco99} 
J.~P.~B.~C.~de~Melo, H.~W.~L.~Naus and 
T.~Frederico, Phys.~Rev.~{\bf C59}, 2278 (1999).


\bibitem{Pacheco2002} J.~P.~B.~C. de Melo, 
T. Frederico, E. Pace and G. Salm\`e, 
Nucl. Phys. {\bf A707}, 399 (2002); ibid. Braz. J. Phys. {\bf 33}, 301 (2003).

\bibitem{Fabiano2007}
Fabiano~P.~Pereira, J.~P.~B.~C.~de~Melo, T.~Frederico, 
Lauro Tomio, Nucl.Phys. {\bf A610},~610~(2007).
  
\bibitem{Pacheco2006} 
  J.~P.~B.~C.~de Melo, T.~Frederico, E.~Pace and G.~Salm\`e,
  Phys.~Lett.~{\bf B581}, 75 (2004); ibid.  Phys.\ Rev.\ D {\bf 73}, 074013 (2006).
  
\bibitem{Leitner2011} O.~Leitner, 
J.-F.~Mathiot, N.~A.~Tsirova,~Eur.Phys.J. {\bf A47} (2011), 17.

\bibitem{Edson2012} Edson O. da Silva, 
J.~P.~B.~C. de Melo, Bruno El-Bennich, Victo S. Filho,
Phy.~Rev.~{\bf C86},~038202~(2012).
  
\bibitem{Pacheco97}J.~P.~B.~C.~de Melo and T.~Frederico, 
Phy. Rev.~{\bf C55},~2043~(1997). 

\bibitem{Hawes1999}F. T. Hawes and M. A. Pichowsky, Phy. Rev. {\bf C55}, 2638 (1999).
    
\bibitem{Tomasi2007} C. Adamuscin, G. I. Gakh and E. Tomasi-Gustafsson, 
Phys. Rev. {\bf C75}, 065201 (207).
    
\bibitem{Wang2007} Z.G.Wang and S.L.Wan, Phy. Rev. {\bf C76}, 025207 (2007).

\bibitem{Braguta2010}V.~V.~Braguta and A.~I.~Onishchenko,
 Phy. Rev. {\bf D70}, 033001 (2004).

\bibitem{Tobias2010} T.~Frederico, E.~Pace, Silva Pisano, G.~Salm\'e, 
Nucl.~Phys,~{\bf B199},~270~(2010).

\bibitem{Maris2008} M.~S.~Bhagwat and P.~Maris,
 Phy. Rev. {\bf C77}, 025203 (2008).

\bibitem{Choi2011}
  H.~M.~Choi and C.~R.~Ji,
  Phys.\ Lett.\  B {\bf 696} (2011) 518.


\bibitem{Pacheco2004}J.~P.~B.~C.~de Melo and T.~Frederico,Brazilian Journal 
of Physics,~{\bf Vol.~34, no.~2A},~881~(2004).

\bibitem{Pacheco2012}J. P. B. C. de Melo and T. Frederico, 
Phy. Lett.,{\bf B708},~87~(2012).

\bibitem{Roberts2013}M.~Pitschmann, C.-Y. Seng, M.~J.~R.-Musolf, 
C. D. Roberts, S.~M.~Schmidt and D.~J. Wilson, 
Phys.~Rev.{\bf C87}, 015205~(2013).

\bibitem{Cardarelli95} F.~Cardarelli,I.L. Grach, 
I. M. Narodetskii, E.~Pace, G. Salm\'e, S. Simula, 
Phy. Lett. {\bf B349},~393~(1995).

\bibitem{Inna84} I.L.Grach and  L.A.Kondratyuk,
Sov. J. Nucl. Phys. {\bf 38}, 198 (1984), ibid., 
I.L.Grach, L.A. Kondratyuk, and M.Strikman,
Phys. Rev. Lett. {\bf 62}, 387 (1989).

\bibitem{Hiller92} S.J.Brodsky and J.Hiller, Phy.~Rev.{\bf D46}, (1992) 2141.

\bibitem{Ji2002}B.L.G. Bakker and C.R. Ji, 
Phy. Rev. D{\bf 65},~116001~(2002).

\bibitem{Choi2004}Ho-Meoyng Choi and C.R. Ji,
Phy. Rev.{\bf D70}, 053015 (2004).



\end{thebibliography}
\end{document}